\documentclass[11pt]{article}
\usepackage{amssymb,latexsym,amsmath,amsbsy,amsthm}
\usepackage[dvips]{graphicx}
\usepackage{xcolor}
\usepackage{cite}
\usepackage{stmaryrd}  

\newtheorem{theo}{Theorem}
\newtheorem{defi}[theo]{Definition}

\newtheorem{prop}[theo]{Proposition}

\def\nn{\nonumber}
\def\deg{\mathop{\rm deg}\nolimits}

\def\qdots{\mathinner{\mkern1mu\raise1pt\vbox{\kern7pt\hbox{.}}\mkern2mu \raise4pt\hbox{.}\mkern2mu\raise7pt\hbox{.}\mkern1mu}}
\def\Z{{\mathbb Z}}
\def\N{{\mathbb N}}

\def\gl{\mathfrak{gl}}
\def\ssl{\mathfrak{sl}}

\def\g{\mathfrak{g}}

\def\so{\mathfrak{so}}

\def\osp{\mathfrak{osp}}
\def\pso{\mathfrak{pso}}
\def\lb{\llbracket}
\def\rb{\rrbracket}

\DeclareMathOperator{\Str}{Str}
\DeclareMathOperator{\tr}{tr}

\def\ra{\rangle}
\def\la{\langle}
\begin{document}
\begin{center}
{\Large \bf
A class of representations of the $\Z_2\times\Z_2$-graded special linear \\[3mm] Lie superalgebra 
$\ssl (m_1+1,m_2|n_1,n_2)$ and quantum statistics} \\[5mm]
{\bf N.I.~Stoilova}\footnote{E-mail: stoilova@inrne.bas.bg}\\[1mm] 
Institute for Nuclear Research and Nuclear Energy, Bulgarian Academy of Sciencies,\\ 
Boul.\ Tsarigradsko Chaussee 72, 1784 Sofia, Bulgaria\\[2mm] 
{\bf J.\ Van der Jeugt}\footnote{E-mail: Joris.VanderJeugt@UGent.be}\\[1mm]
Department of Mathematics, Computer Science and Statistics, Ghent University,\\
Krijgslaan 281-S9, B-9000 Gent, Belgium.
\end{center}

\vskip 2 cm

\begin{abstract}
\noindent 
The description of the $\Z_2\times\Z_2$-graded special linear Lie superalgebra 
$\ssl (m_1+1,m_2|n_1,n_2)$ is carried out via generators $a_1^\pm,\ldots, a_{m_1+m_2+n_1+n_2}^\pm$
that satisfy triple relations and are
called creation and annihilation operators. With respect to these generators, a class of 
Fock type representations  of $\ssl (m_1+1,m_2|n_1,n_2)$ is constructed. The properties of the
underlying statistics are discussed and  its Pauli principle is formulated.
\end{abstract}

\setcounter{equation}{0}
\section{Introduction} \label{sec:A}%

A result from quantum field theory is that particles with half-integer spins are fermions,
satisfying the Fermi-Dirac statistics, and particles with integer spins are bosons, satisfying 
the Bose-Einstein statistics. However, beyond Fermi-Dirac and Bose-Einstein statistics, 
various kinds of generalized quantum statistics have been 
introduced, investigated and discussed~(see e.g. \cite{ Sanchez} for a review and references. 
One of the first such generalizations are the so called parafermion and paraboson statistics~\cite{Green}. The algebraic 
structure behind a system of $2m$-parafermion operators is the orthogonal Lie algebra $\so(2m+1)$~\cite{KR, RS},
and behind a system of $2n$-parabosons is the orthosymplectic Lie superalgebra $\osp(1|2n)$~\cite{Ganchev}.
For a mixed system of $2m$-parafermions and $2n$-parabosons there are two types of mutual nontrivial relations
(from a physical point of view)~\cite{Greenberg} with algebraic structures the Lie superalgebra 
$\osp(2m+1|2n)$~\cite{Palev1} 
and the $\Z_2 \times \Z_2$-graded Lie superalgebra $\osp(1,2m|2n,0)\equiv \pso(2m+1|2n)$~\cite{Tolstoy2014, SV2018}.
All these algebras are of type $B$ Lie algebras, Lie superalgebras or  $\Z_2 \times \Z_2$-graded Lie superalgebras.
Furthermore, generalized statistics have been associated with all classical Lie algebras and basic classical
Lie superalgebras from the infinite series $A, B, C,$ and $D$ and are refered to as $A, B, C$ 
and $D$-(super)statistics~\cite{Palev_97, SJLA, SJLSA}.
Therefore it is natural to consider their $\Z_2 \times \Z_2$-graded counterparts. 

In recent years there has been an increased interests in colour algebras introduced in~\cite{Rit1,Rit2,Scheunert1979}, especially in those with 
$\Z_2 \times \Z_2$-grading and their applications. In the following section we will remind the concept of 
 $\Z_2 \times \Z_2$-graded Lie superalgebras, and define the 
$\Z_2 \times \Z_2$-graded special linear Lie superalgebra $\ssl(m_1+1,m_2|n_1,n_2)$. In particular we shall give a set 
of $\ssl(m_1+1,m_2|n_1,n_2)$ generators satisfying triple relations. Section 3 is devoted to a class 
of $\ssl(m_1+1,m_2|n_1,n_2)$ irreducible representations, the so called Fock representations and in Section 4 we discuss the 
$\Z_2 \times \Z_2$-graded $A$ statistics.

\section{$\Z_2\times\Z_2$-graded special linear Lie superalgebra $\ssl(m_1+1,m_2|n_1,n_2)$}
\setcounter{equation}{0} \label{sec:B}

First it will be useful to  recall the definition of $\Z_2\times\Z_2$-graded 
Lie superalgebras~\cite{Rit1,Rit2,Scheunert1979}.

As a linear space such algebras $\g$ are direct sums of four subspaces:
\begin{equation}
\g=\bigoplus_{\boldsymbol{a}} \g_{\boldsymbol{a}} =
\g_{(0,0)} \oplus \g_{(0,1)} \oplus \g_{(1,0)} \oplus \g_{(1,1)} 
\label{ZZgrading}
\end{equation}
with $\boldsymbol{a}=(a_1,a_2)$  an element of $\Z_2\times\Z_2$.
Writting $x_{\boldsymbol{a}}, y_{\boldsymbol{a}},\ldots$,
means that these elements belong to $\g_{\boldsymbol{a}}$  and $\boldsymbol{a}$ is said to be the degree, $\deg x_{\boldsymbol{a}}$, of $x_{\boldsymbol{a}}$.
Such elements $x_{\boldsymbol{a}}$ are called homogeneous elements.
The bracket $\lb\cdot,\cdot\rb$ on a $\Z_2\times\Z_2$-graded Lie superalgebra $\g$ 
satisfies the grading, symmetry and Jacobi identities:
\begin{align}
& \lb x_{\boldsymbol{a}}, y_{\boldsymbol{b}} \rb \in \g_{\boldsymbol{a}+\boldsymbol{b}}, \label{grading}\\
& \lb x_{\boldsymbol{a}}, y_{\boldsymbol{b}} \rb = -(-1)^{\boldsymbol{a}\cdot\boldsymbol{b}} 
\lb y_{\boldsymbol{b}}, x_{\boldsymbol{a}} \rb, \label{symmetry}\\
& \lb x_{\boldsymbol{a}}, \lb y_{\boldsymbol{b}}, z_{\boldsymbol{c}}\rb \rb =
\lb \lb x_{\boldsymbol{a}}, y_{\boldsymbol{b}}\rb , z_{\boldsymbol{c}} \rb +
(-1)^{\boldsymbol{a}\cdot\boldsymbol{b}} \lb y_{\boldsymbol{b}}, \lb x_{\boldsymbol{a}}, z_{\boldsymbol{c}}\rb \rb,
\label{jacobi}
\end{align} 
where
\begin{equation}
\boldsymbol{a}+\boldsymbol{b}=(a_1+b_1,a_2+b_2)\in \Z_2\times\Z_2, \qquad
\boldsymbol{a}\cdot\boldsymbol{b} = a_1b_1+a_2b_2.
\label{sign}
\end{equation}
 From~\eqref{grading} and~\eqref{sign} it follows that $\g_{(0,0)}$ is a Lie subalgebra of $\g$, and 
$\g_{(1,0)}, \g_{(0,1)}, \g_{(1,1)}$ are $\g_{(0,0)}$-modules. In addition $\g_{(0,0)}\oplus \g_{(1,1)}$
is also a Lie subalgebra of $\g$, and the subspace $\g_{(1,0)}\oplus \g_{(0,1)}$
 is a $\g_{(0,0)}\oplus \g_{(1,1)}$-module. Furthermore, $\{ \g_{(1,1)}, \g_{(0,1)}\} \subset \g_{(1,0)} $
and $\{ \g_{(1,1)}, \g_{(1,0)}\} \subset \g_{(0,1)} $.

Let $M$ be an arbitrary $(m_1+m_2+n_1+n_2+1  \times m_1+m_2+n_1+n_2+1)$-matrix of the following block form:
\begin{equation}
M = \begin{array}{c c}
    {\begin{array} {@{} c c  cc @{}}  m_1+1 & m_2\ & \ n_1\ & \ \ n_2 \ \end{array} } & {} \\  
    \left(\begin{array}{cccc} 
a_{(0,0)} & a_{(1,1)} & a_{(1,0)} & a_{(0,1)} \\[1mm] 
b_{(1,1)} & b_{(0,0)} & b_{(0,1)} & b_{(1,0)} \\[1mm] 
c_{(1,0)} & c_{(0,1)} & c_{(0,0)} & c_{(1,1)} \\[1mm] 
d_{(0,1)} & d_{(1,0)} & d_{(1,1)} & d_{(0,0)} 
    \end{array}\right)
 & \hspace{-2mm} 
		{\begin{array}{l}
     m_1+1 \\[1mm]  m_2 \\[1mm]	n_1 \\[1mm] n_2
    \end{array} } \\ 
  \end{array} 
\label{ZZsl}
\end{equation}
The indices of the matrix blocks show  the $\Z_2\times\Z_2$-grading, and the size of the blocks is indicated in the lines above and to the right of the matrix. Matrix $M$ can be written as a sum of four matrices: 
\begin{equation}
M=M_{(0,0)}+M_{(1,1)}+M_{(1,0)}+M_{(0,1)}, \label{4M}
\end{equation}
with $M_{(a,b)}$ the corresponding block matrices $m_{(a,b)}$ and all other zeros.
Defining the bracket $\lb\cdot,\cdot\rb$ on the space of these matrices by:

\begin{equation}
\lb M_{(a_1,a_2)}, \tilde{M}_{(b_1,b_2)}\rb = 
M_{(a_1,a_2)}\tilde{M}_{(b_1,b_2)} -(-1)^{a_1b_1+a_2b_2}\tilde{M}_{(b_1,b_2)}M_{(a_1,a_2)}
\label{BracketH}
\end{equation}
for the homogeneous elements $M_{(a_1,a_2)}$ and $\tilde{M}_{(b_1,b_2)}$ and extending it by linearity 
one obtains the $\Z_2\times\Z_2$-graded general linear Lie superalgebra $\gl(m_1+1, m_2|n_1,n_2)$.
 
It is straightforward to check that for the graded supertrace  $\Str(A)=\tr(a_{(0,0)})+\tr(b_{(0,0)})-\tr(c_{(0,0)})-\tr(d_{(0,0)})$  in terms of the ordinary trace $\tr$,  $\Str \lb A,B \rb =0$. 
Therefore,  the $\Z_2\times\Z_2$-graded special linear Lie superalgebra $\ssl(m_1+1,m_2|n_1,n_2)$ is defined as the subalgebra  of $\gl(m_1+1,m_2|n_1,n_2)$ with graded supertrace equal to~0.

Let
\begin{align}
d_i &= 
\left\{  
\begin{array}{rl}
	(0,0); & i=0,\ldots,m_1\\
	(1,1); & i=m_1+1,\ldots,m_1+m_2=m\\
	(1,0); & i=m_1+m_2+1,\ldots,m_1+m_2+n_1=m+n_1\\
	(0,1); & i=m_1+m_2+n_1+1,\ldots,m_1+m_2+n_1+n_2=m+n,
\end{array}
\right.\label{di}
\end{align}
and let $e_{ij}, \; i,j=0,1,\ldots,m_1+m_2+n_1+n_2=m+n \;$ (where $m_1+m_2=m, n_1+n_2=n) $ be the $(m+n+1 \times m+n+1)$ 
matrix~\eqref{ZZsl} with $1$ in the entry of row $i$, column $j$ and $0$ elsewhere. These matrices are homogeneous  
and the grading $\deg(e_{ij})$ is as follows:
\begin{equation}
\deg(e_{ij})=d_i+d_j.
\nn
\end{equation} 
The bracket for such matrices is given by:
\begin{equation}
\lb e_{ij}, e_{kl}\rb = \delta_{jk}e_{il}-(-1)^{(d_i+d_j)\cdot(d_k+d_l)}\delta_{il}e_{kj}.
\nn
\end{equation} 
The algebra $\ssl(m_1+1,m_2|n_1,n_2)$ can be considered as the linear envelope of 
$e_{ij}, \; i\neq j=0,1,\ldots,m+n, \; e_{00}-(-1)^{d_i\cdot d_i}e_{ii}, \; i=1,\ldots,m+n$.
A set of generators of $\ssl(m_1+1,m_2|n_1,n_2)$ is given by:
\begin{equation}
a_i^+=e_{i0},\;\; a_i^-=e_{0i},\;i=1,\ldots,m+n, \;\; (\deg(a_i^{\pm})=d_i)
\end{equation}
since
\begin{equation}
\lb a_i^+, a_j^- \rb =e_{ij},\;i\neq j=1,\ldots,m+n;\; 
\lb a_k^-, a_k^+\rb =e_{00}-(-1)^{d_k\cdot d_k}e_{kk},\;k=1,\ldots,m+n.
\end{equation}
Denote these generators by
\begin{align}
& a_i^{\pm}\equiv b_i^{\pm}\in \g_{(0,0)}, \;\;i=1,\ldots,m_1, \label{1bi}\\
& a_i^{\pm}\equiv \tilde{b}_{i-m_1}^{\pm}\in \g_{(1,1)}, \;\;i=m_1+1,\ldots,m, \label{2bi}\\
& a_i^{\pm}\equiv {f}_{i-m}^{\pm}\in \g_{(1,0)}, \;\;i=m+1,\ldots,m+n_1, \label{1fi}\\
& a_i^{\pm}\equiv \tilde{f}_{i-m-n_1}^{\pm}\in \g_{(0,1)}, \;\;i=m+n_1+1,\ldots,m+n \label{2fi}.
\end{align}
The $\Z_2\times\Z_2$-graded special linear Lie superalgebra $\ssl(m_1+1,m_2|n_1,n_2)$
can be defined in terms of the generators $a_i^{\pm},\; i=1,\ldots, m+n$  and the following relations:
\begin{align}
& \lb a_i^{\xi}, a_j^{\xi} \rb =0, \;\;\xi=\pm, \;i,j=1,\ldots,m+n, \nn\\
& \lb \lb a_i^{+}, a_j^{-} \rb, a_k^+\rb =\delta_{jk}a_i^++(-1)^{d_i\cdot d_i}\delta_{ij}a_k^+, \label{Rel}\\
& \lb \lb a_i^{+}, a_j^{-} \rb, a_k^-\rb =-(-1)^{(d_i+d_j)\cdot d_k}\delta_{ik}a_j^--(-1)^{d_i\cdot d_i}\delta_{ij}a_k^-, 
\;\;i,j,k=1,\ldots,m+n. \nn
\end{align}

\begin{defi}
We will call the generators $a_i^\pm,\; i=1,\ldots,m+n$, creation and annihilation operators of the 
$\Z_2\times\Z_2$-graded Lie superalgebra $\ssl(m_1+1,m_2|n_1,n_2)$.
\end{defi}

\setcounter{equation}{0}
\section{Representations of the $\Z_2\times\Z_2$-graded Lie superalgebra $\ssl(m_1+1,m_2|n_1,n_2)$} 
\label{sec:C}%
We now proceed to construct a class of representations, Fock type representations, of the 
$\Z_2\times\Z_2$-graded Lie superalgebra $\ssl(m_1+1,m_2|n_1,n_2)$.
The irreducible Fock representations are
labelled by one non-negative integer $p=1,2,\ldots$, 
called an order of the statistics. To construct them assume that the
corresponding representation space $W_p$ contains (up to
a multiple) a cyclic vector $|0\ra$, such that
\begin{align}
& a_i^-|0\ra =0,\quad i=1,2,\ldots ,n+m,\nn\\
& \lb  a_{ i}^-,a_{ j}^+\rb |0\ra=\delta_{ij} p|0\ra,\quad p\in \N,\quad i,j=1,2,\ldots ,n+m \label{Fock}.
\end{align}
The Fock spaces are finite-dimensional irreducible $\ssl(m_1+1,m_2|n_1,n_2)$-modules. The vectors
\begin{equation}
(b_1^+)^{r_1}\ldots (b_{m_1}^+)^{r_{m_1}}(\tilde{b}_1^+)^{l_1}\ldots (\tilde{b}_{m_2}^+)^{l_{m_2}}
(f_1^+)^{\theta_1}\ldots (f_{n_1}^+)^{\theta_{n_1}}(\tilde{f}_1^+)^{\lambda_1}\ldots (\tilde{f}_{n_2}^+)^{\lambda_{n_2}}
|0\ra \label{BasisInfinity}
\end{equation}
subject to the following restriction
\begin{equation}
r_i, l_i\in \Z_+,\;\; \quad \theta_{i}, \lambda_i\in \{0,1\}, \;\;
R=\sum_{i=1}^{m_1}r_i+\sum_{i=1}^{m_2}l_i+\sum_{i=1}^{n_1}\theta_i+\sum_{i=1}^{n_2}\lambda_i \leq p \label{Conditions1}
\end{equation}
constitute a basis in $W_p$.
The linear space of all vectors~\eqref{BasisInfinity} for any $r_i, l_i\in \Z_+,\; \theta_{i}, \lambda_i\in \{0,1\}$,
i.e. without the restriction~\eqref{Conditions1}, is an infinite-dimensional $\ssl(m_1+1,m_2|n_1,n_2)$-module $\bar{W}_p$. 
It is however not irreducible and $\bar{W}_p$ contains an infinite-dimensional invariant subspace $W_p^{inv}$, linear envelope 
of all the vectors~\eqref{BasisInfinity} with 
$R=\sum_{i=1}^{m_1}r_i+\sum_{i=1}^{m_2}l_i+\sum_{i=1}^{n_1}\theta_i+\sum_{i=1}^{n_2}\lambda_i > p$. Then 
$W_p$ is a factor module of $\bar{W}_p$ with respect to $W_p^{inv}$.

Define a Hermitian form $\langle\,,\,\rangle$ on $W_p$  with the standard Fock space technique. So, postulate that

\begin{align}
& \la 0| 0\ra  =1,\\
& \la a_i^\pm v | w \ra = \la v | a_i^\mp w \ra, \qquad\forall v,w\in W_p.
\end{align}
With respect to this form, the different vectors in~\eqref{BasisInfinity} 
are orthogonal, and the following vectors form an orthonormal
basis of $W_p$~:

\begin{align}
& |p;r_1,\ldots,r_{m_1},l_1,\ldots,l_{m_2},\theta_1,\ldots,\theta_{n_1},\lambda_1,\ldots,\lambda_{n_2})=\nn
\sqrt{\frac{(p-R)!}
 {p!r_1!\ldots \lambda_{n_2}!}}\nn\\
&\nn\\
&
\times (b_1^+)^{r_1}\ldots (b_{m_1}^+)^{r_{m_1}}(\tilde{b}_1^+)^{l_1}\ldots (\tilde{b}_{m_2}^+)^{l_{m_2}}
(f_1^+)^{\theta_1}\ldots (f_{n_1}^+)^{\theta_{n_1}}(\tilde{f}_1^+)^{\lambda_1}\ldots (\tilde{f}_{n_2}^+)^{\lambda_{n_2}}
|0\ra, \label{Vectors}
\end{align}
with
\begin{equation}
r_i, l_i\in \Z_+,\;\; \quad \theta_{i}, \lambda_i\in \{0,1\}, \;\;
R=\sum_{i=1}^{m_1}r_i+\sum_{i=1}^{m_2}l_i+\sum_{i=1}^{n_1}\theta_i+\sum_{i=1}^{n_2}\lambda_i \leq p. \label{Conditions}
\end{equation}

\begin{prop} 
The transformation of the orthonormal  basis of $W_p$ under the
action of the creation and annihilation operators $a_i^{\pm}$  reads: 
\begin{align}
&b_i^+|p;\ldots,r_{i-1},r_i,r_{i+1},\ldots )=
\sqrt{(r_i+1)(p-R)}
|p;\ldots,r_{i-1},r_i+1,r_{i+1},\ldots ), \label{b+}\\
&\nn\\
&\tilde{b}_i^+|p;\ldots,l_{i-1},l_i,l_{i+1},\ldots )=
\sqrt{(l_i+1)(p-R)}
|p;\ldots,l_{i-1},l_i+1,l_{i+1},\ldots ), \label{tilde_b+}\\
&\nn\\
&f_i^+|p;\ldots,\theta_{i-1},\theta_i,\theta_{i+1},\ldots )=(1-\theta_i)(-1)^{l_1+\ldots+ l_{m_2}}
(-1)^{\theta_1+\ldots + \theta_{i-1}}
\sqrt{p-R}
|p;\ldots,\theta_{i-1},\theta_i+1,\theta_{i+1},\ldots ), \label{f+}\\
&\tilde{f}_i^+|p;\ldots,\lambda_{i-1},\lambda_i,\lambda_{i+1},\ldots )=(1-\lambda_i)(-1)^{l_1+\ldots+ l_{m_2}}
(-1)^{\lambda_1+\ldots + \lambda_{i-1}}
\sqrt{p-R}
|p;\ldots,\theta_{i-1},\theta_i+1,\theta_{i+1},\ldots ) \label{tilde_f+},
\end{align}
\begin{align}
&b_i^-|p;\ldots,r_{i-1},r_i,r_{i+1},\ldots )=
\sqrt{r_i(p-R+1)}
|p;\ldots,r_{i-1},r_i-1,r_{i+1},\ldots ), \label{b-}\\
&\nn\\
&\tilde{b}_i^-|p;\ldots,l_{i-1},l_i,l_{i+1},\ldots )=
\sqrt{l_i(p-R+1)}
|p;\ldots,l_{i-1},l_i-1,l_{i+1},\ldots ), \label{tilde_b-}\\
&\nn\\
&f_i^-|p;\ldots,\theta_{i-1},\theta_i,\theta_{i+1},\ldots )=\theta_i(-1)^{l_1+\ldots+ l_{m_2}}
(-1)^{\theta_1+\ldots + \theta_{i-1}}
\sqrt{p-R+1}
|p;\ldots,\theta_{i-1},\theta_i-1,\theta_{i+1},\ldots ), \label{f-}\\
&\tilde{f}_i^-|p;\ldots,\lambda_{i-1},\lambda_i,\lambda_{i+1},\ldots )=\lambda_i(-1)^{l_1+\ldots+ l_{m_2}}
(-1)^{\lambda_1+\ldots + \lambda_{i-1}}
\sqrt{p-R+1}
|p;\ldots,\theta_{i-1},\theta_i+1,\theta_{i+1},\ldots ) \label{tilde_f-}.
\end{align}
\end{prop}
\noindent
The above transformations are obtained by applying the defining relations~\eqref{Rel} of $\ssl(m_1+1,m_2|n_1,n_2)$.
However, in order to prove that these explicit actions~\eqref{b+}-\eqref{tilde_f-} give a representation of the 
$\Z_2\times\Z_2$-graded special linear Lie superalgebra $\ssl(m_1+1,m_2|n_1,n_2)$ it is sufficient to show 
that~\eqref{b+}-\eqref{tilde_f-} satisfy the defining relations of the algebra which is a long but 
straightforward matter. The irreducibility then follows from the fact that for any nonzero vector 
$x\in W_p$ there exists a polynomial $\mathfrak{P}$ of $\ssl(m_1+1,m_2|n_1,n_2)$ generators such that 
$\mathfrak{P}x=W_p$.

\setcounter{equation}{0} 
\section{ $\Z_2\times\Z_2$-graded $A$  statistics} 
\label{sec:D}%

The operators $b_i^\pm\equiv a_i^{\pm}\in \g_{(0,0)}, \;\; i=1,\ldots,m_1$ and 
$\tilde{b}_{i-m_1}^\pm\equiv a_i^{\pm}\in \g_{(1,1)}, \;\; i=m_1+1,\ldots,m$ generate the Lie algebra $\ssl(m+1)$
and satisfy the relations:
\begin{align}
& [ a_i^{\xi}, a_j^{\xi} ] =0, \;\;\xi=\pm, \;i,j=1,\ldots,m, \nn\\
& [\; [ a_i^{+}, a_j^{-} ], a_k^+] =\delta_{jk}a_i^++\delta_{ij}a_k^+, \label{A-stat}\\
& [\; [ a_i^{+}, a_j^{-} ], a_k^-] =-\delta_{ik}a_j^--\delta_{ij}a_k^-. \nn
\end{align}
Relations~\eqref{A-stat} are the defining triple relations of $A$-statistics~\cite{Jellal}.

There are other two sets of operators $f_{i-m}^{\pm}\equiv a_i^{\pm}\in \g_{(1,0)}$ for $i=m+1,\ldots, m+n_1$ 
and $\tilde{f}_{i-m-n_1}^{\pm}\equiv a_i^{\pm}\in \g_{(0,1)}$ for $i=m+n_1+1,\ldots, m+n$ satisfying the 
common defining triple relations of $A$-superstatistics~\cite{Palev2003}: 

\begin{align}
& \{ a_i^+, a_j^+ \} = \{ a_i^-, a_j^- \} = 0 , \nonumber \\
& [ \{ a_i^+, a_j^- \}, a_k^+ ] = \delta_{jk} a_i^+ - \delta_{ij} a_k^+,\nonumber \\
& [ \{ a_i^+, a_j^- \}, a_k^- ] = -\delta_{ik} a_j^- + \delta_{ij} a_k^- . \label{A1}
\end{align}
Here, either $i,j,k= m+1,\ldots, m+n_1$ or else 
$i,j,k= m+n_1+1,\ldots,m+n$.

It is easy to write down the mixed relations between the three families of operators $a_i^\pm, \;i=1,\ldots, m+n$
(the operators of $A$-statistics and the two sets of operators of $A$-superstatistics),
in terms of  (anti)commutators using~\eqref{Rel} and the $\Z_2\times\Z_2$-grading.
In particular  the mixed relations between the two families of $A$-superstatistics operators are as follows:
\begin{align}
& [ a_i^+, a_j^+ ] = [ a_i^-, a_j^- ] = 0 , \nonumber \\
& \{ [a_i^+, a_j^- ], a_k^+ \} = \delta_{jk} a_i^+, \nonumber\\
& \{ [ a_i^+, a_j^- ], a_k^- \} = \delta_{ik} a_j^- , \label{MixA1}
\end{align}
with in~\eqref{MixA1}, either $i= m+1,\ldots,m+n_1$, $j=m+n_1+1,\ldots,m+n$, $k=m+1,\ldots,m+n$,
or else $i=m+n_1+1,\ldots,m+n$, $j=m+1\ldots,m+n_1$, $k=m+1,\ldots,m+n$;

and 
 \begin{align}
& \{ a_i^+, a_j^+ \} = \{ a_i^-, a_j^- \} = 0 , \nonumber \\
& [ \{ a_i^+, a_j^- \}, a_k^+ ] =  - \delta_{ij} a_k^+,\nonumber \\
& [ \{ a_i^+, a_j^- \}, a_k^- ] =  + \delta_{ij} a_k^- . \label{MixA2}
\end{align}
with in~\eqref{MixA2}, either $i,j=m+1,\ldots,m+n_1$, $k=m+n_1+1,\ldots,m+n$,
or else $i,j = m+n_1+1,\ldots,m+n$, $k=m+1,\ldots,m+n_1$.

The operators $a_i^+$ (resp.  $a_i^-$) can be interpreted as 
operators in a state space $W_p$, the Fock space, creating  ``a particle"   (resp.  annihilating ``a particle"), with
energy $\varepsilon_i$ . Let for simplicity 
$m=n$ and 
consider a  Hamiltonian 
\begin{equation}
H=
\sum_{i=1}^{m}\varepsilon_i ([a_i^+,a_i^-]+\{a_i^+,a_i^-\}).
\label{H}
\end{equation}

Then 
\begin{equation}
[H,a_i^\pm]=\pm \varepsilon_i a_i^\pm,~~~
[H,a_{i+m}^\pm]=\pm \varepsilon_i a_{i+m}^\pm. \label{rel_H_ai} 
\end{equation}
This result together with~\eqref{b+}-\eqref{tilde_f-}  allows one 
to interpret $r_i, \;i=1,\ldots, m_1$, $l_i, \; i=1,\ldots, m_2$, $\theta_i, \; i=1,\ldots, n_1$ and 
$\lambda_i,\; i=1,\dots, n_2$ as the number of particles on the corresponding orbital.
The operator $a_i^+$  increases this number by
one, it ``creates'' a particle in the one-particle state (orbital).  Similarly, the operator $a_i^-$  diminishes 
this number by one, it ``kills'' a particle on the corresponding orbital. However on
every orbital of the last $n$ there cannot be more than one particle,
whereas such restriction does not hold for the first $m$ orbitals.
Therefore we have generalized fermions and bosons in this case. However there is an extra property.
Since $\sum_{i=1}^{m_1}r_i+\sum_{i=1}^{m_2}l_i+\sum_{i=1}^{n_1}\theta_i+\sum_{i=1}^{n_2}\lambda_i \leq p$
no more than $p$ particles can be accomodated in the system. This is the {\it Pauli principle} for this statistics.

\section{Concluding Remarks}\label{sec:5}

We defined the $\Z_2\times\Z_2$-graded Lie superalgebra $\mathfrak{sl}(m_1+1,m_2|n_1,n_2)$ 
in terms of a set of  generators and triple relations. A class of Fock type $\mathfrak{sl}(m_1+1,m_2|n_1,n_2)$ representations is constructed under the condition 
\begin{equation}
(a_i^+)^{\dag} = a_i^-, \quad i=1,\ldots, m_1+m_2+n_1+n_2 \nonumber
\end{equation}
motivated by the physical requirement that physical observables be represented by Hermitian operators. This constraint necessarily implies that the generalized bosons act in finite-dimensional state spaces. The restriction that no more than $p$ particles ($p$ – the order of the statistics) may occupy the system at a given time situates the 
$\Z_2\times\Z_2$-graded $A$ statistics considered here within the broader category of so called exclusion statistics. 

In the present study, the generalized $\Z_2\times\Z_2$-graded $A$ quantum statistics has been introduced primarily through a mathematical framework, focusing on the underlying algebraic structure. A natural and promising direction for future research is the application of this statistical formalism to the analysis of concrete physical systems and phenomena — in particular, to systems exhibiting strongly correlated behavior, such as those encountered in the quantum Hall effect.

\section*{Acknowledgments}

This article is based upon work from COST Action 21109 CaLISTA, supported by COST 
(European Cooperation in Science and Technology) and 
by the Bulgarian National Science Fund, grant KP-06-N88/3.

\end{document}